\documentstyle[preprint,aps]{revtex}

\begin{document}
\draft
\title{{\bf Fluctuations and Transients in Quantum-Resonant Evolution}}
\author{{\bf Itzhack Dana}$^{1}${\bf \ and Dmitry L. Dorofeev}$^{2}$}
\address{$^{1}$Minerva Center and Department of Physics, Bar-Ilan University,
Ramat-Gan 52900, Israel}
\address{$^{2}$Department of Physics, Voronezh State University, Voronezh 394693,
Russia}
\maketitle

\begin{abstract}
The quantum-resonant evolution of the mean kinetic energy (MKE) of the
kicked particle is studied in detail on different time scales for {\em 
general} kicking potentials. It is shown that the asymptotic time behavior
of a wave-packet MKE is typically a linear growth with bounded fluctuations
having a simple number-theoretical origin. For a large class of wave
packets, the MKE is shown to be exactly the superposition of its asymptotic
behavior and transient logarithmic corrections. Both fluctuations and
transients can be significant for not too large times but they may vanish
identically under some conditions. In the case of incoherent mixtures of
plane waves, it is shown that the MKE never exhibits asymptotic fluctuations
but transients usually occur.\newline
\end{abstract}


The recent extensive studies of quantum resonance (QR) and related phenomena
exhibited by the periodically kicked particle (KP), either in the presence
or in the absence of gravity, have led to interesting new connections
between the classical and quantum dynamics of nonintegrable systems \cite
{ao,ao1,kp1,kp2}. These phenomena are now recognized as highly efficient
methods for imparting large changes in the kinetic energy of cold atoms in
atom-optics experiments \cite{ao,ao1}. The quantum KP in the absence of
gravity is described by the Hamiltonian 
\begin{equation}
\hat{H}=\frac{\hat{p}^{2}}{2}+kV(\hat{x})\sum_{t}\delta (t^{\prime }-t\tau ),
\label{H}
\end{equation}
where $(\hat{x},\ \hat{p})$ are position and momentum operators, $k$ is the
nonintegrability parameter, $V(x)$ is a periodic potential, $t^{\prime }$
and $t$ are the continuous and ``integer'' times, and $\tau $ is the kicking
period; the units are chosen so that the particle mass is $1$, $\hbar =1$,
and the period of $V(x)$ is $2\pi $. The most investigated case of QR, which
is also the easiest to observe experimentally, corresponds to integer values
of $\tau /(2\pi )$. It was shown \cite{kp1,kp2,dd} that in this case QR
manifests itself in an {\em asymptotic linear} growth, as a function of time 
$t$, of the mean kinetic energy (MKE) of either a general KP wave packet or
an incoherent mixture of plane waves.\newline

Experimental observations of the QR phenomenon are, of course, limited in
time. In addition, since integer values of $\tau /(2\pi )$ cannot be
precisely realized experimentally, the growth of the MKE may stop or
saturate after some finite time. This was shown in Ref. \cite{kp2} when
studying the evolution of an incoherent mixture of plane waves for $\tau
=2\pi +\epsilon $ ($|\epsilon |\ll 1$), using the standard potential $
V(x)=\cos (x)$. It was found that the MKE $\left\langle E_{t,\epsilon
}\right\rangle $ of this mixture essentially coincides with that for $
\epsilon =0$ in the time interval $0<t<|k\epsilon |^{-1/2}$, where $
\left\langle E_{t,\epsilon }\right\rangle \approx \left\langle
E_{t,0}\right\rangle $ is already approximately given by the asymptotic
linear behavior of $\left\langle E_{t,0}\right\rangle $; however, for $
t>|k\epsilon |^{-1/2}$, the growth of $\left\langle E_{t,\epsilon
}\right\rangle $ is suppressed due to dynamical localization. In Ref. 
\cite{dd}, the momentum probability distribution of a wave packet for a
two-harmonic potential was shown to be also robust, on some initial time
interval, under small deviations $\epsilon $ of $\tau $ around $\tau =2\pi$.
These facts motivate the investigation of quantum-resonant evolution on 
{\em arbitrary} time scales, not just on asymptotic ones. In particular,
since the asymptotic behavior may start in practice at times $t$ not too
long, it is important to inquire about the exact nature of this behavior,
i.e., whether {\em fluctuations} take place around the (average) linear
growth for {\em general} potentials $V(x)$ and what is the strength of these
fluctuations.\newline

In this paper, the quantum-resonant evolution of the KP is studied in detail
on different time scales for integer $\tau /(2\pi )$ and general $V(x)$. In
the case of the MKE $\left\langle E\right\rangle _{t}$ of wave packets, we
show that asymptotic fluctuations of $\left\langle E\right\rangle _{t}$
occur only if $V(x)$ is {\em multi-harmonic} and the initial wave packet is 
{\em nonuniform} (in a well-defined sense). These fluctuations and the
average linear growth have essentially the same origin, of a simple
number-theoretical nature. The fluctuations have bounded variation in time
and are therefore much smaller than $\left\langle E\right\rangle _{t}$ for
large $t$. For a large class of initial wave packets, we derive exact closed
expressions for $\left\langle E\right\rangle _{t}$: $\left\langle
E\right\rangle _{t}$ is just the superposition of its asymptotic behavior
(linear growth and fluctuations) and transient logarithmic [$\ln (t)$]
corrections. The fluctuations are exactly periodic or quasiperiodic in time
and both they and the transients can be significant for not too large times.
The transients vanish if $V(x)$ and the initial wave packet satisfy some
symmetry conditions. Then, the asymptotic behavior starts at $t=1$. If also
the fluctuations vanish, e.g., for mono-harmonic $V(x)$, $\left\langle
E\right\rangle _{t}$ is an {\em exactly linear} function of $t$. In the case
of incoherent mixtures of plane waves, we show that the MKE never exhibits
asymptotic fluctuations but transients occur for nonuniform mixtures.\newline

{\em Background.} We first summarize known facts \cite{kp1,kp2,dd} about the
KP quantum dynamics. The one-period evolution operator for (\ref{H}), from $
t^{\prime }=t+0$ to $t^{\prime }=t+\tau +0$, is given by 
\begin{equation}
\hat{U}=\exp \left[ -ikV(\hat{x})\right] \exp \left( -i\tau \hat{p}
^{2}/2\right)  \label{U}
\end{equation}
($\hbar =1$). Because of the $2\pi $-periodicity of (\ref{U}) in $\hat{x}$,
it is natural to study the quantum evolution in terms of Bloch functions $
\varphi _{\beta }(x)=\exp (i\beta x)\psi _{\beta }(x)$, where $\beta $ is a
quasimomentum, $0\leq \beta <1$, and $\psi _{\beta }(x+2\pi )=$ $\psi
_{\beta }(x)$. In fact, the application of $\hat{U}$ on $\varphi _{\beta
}(x) $ results in a Bloch function associated with the same quasimomentum, 
$\hat{U}\varphi _{\beta }(x)=\exp (i\beta x)\psi _{\beta }^{\prime }(x)$;
here $\psi _{\beta }^{\prime }(x)$ is the $2\pi $-periodic function $\psi
_{\beta }^{\prime }(x)=\hat{U}_{\beta }\psi _{\beta }(x)$, where 
\begin{equation}
\hat{U}_{\beta }=\exp \left[ -ikV(\hat{x})\right] \exp \left[ -i\tau \left( 
\hat{p}+\beta \right) ^{2}/2\right] .  \label{Ub}
\end{equation}
Since the operator (\ref{Ub}) acts only on $2\pi $-periodic functions $\psi
_{\beta }(x)$, one can interpret $x$ as an angle and $\hat{p}$ as an
angular-momentum operator with integer eigenvalues $n$. Then, (\ref{Ub}) may
be viewed as the one-period evolution operator for a ``$\beta $-kicked
rotor'' ($\beta $-KR). Given an arbitrary KP wave packet $\Psi (x)$ and
using the relation between $\Psi (x)$ and its momentum representation $
\widetilde{\Psi }(p)$, it is easy to see that $\Psi (x)$ can be always
expressed as a superposition of Bloch functions, $\Psi
(x)=\int_{0}^{1}d\beta \exp (i\beta x)\psi _{\beta }(x)$, where 
\begin{equation}
\psi _{\beta }(x)=\frac{1}{\sqrt{2\pi }}\sum_{n}\widetilde{\Psi }(n+\beta
)\exp (inx).  \label{phb}
\end{equation}
One then gets the basic relation 
\begin{equation}
\Psi _{t}(x)\equiv \hat{U}^{t}\Psi (x)=\int_{0}^{1}d\beta \exp (i\beta x)
\hat{U}_{\beta }^{t}\psi _{\beta }(x)  \label{KPR}
\end{equation}
for integer time $t$, connecting the quantum dynamics of the KP with that of 
$\beta $-KRs. The evolution of a $\beta $-KR wave packet $\psi _{\beta }(x)$
under the QR condition $\tau /(2\pi )=l$, where $l$\ is a positive integer,
was studied in Ref.\cite{dd} for general $2\pi $-periodic potential $V(x)$, 
\begin{equation}
V(x)=\sum_{m}V_{m}\exp (-imx).  \label{V}
\end{equation}
It was found that, up to a nonrelevant constant phase factor, one has 
\cite{dd} 
\begin{equation}
\psi _{\beta ,t}(x)\equiv \hat{U}_{\beta }^{t}\psi _{\beta }(x)=\exp \left[
-ik\bar{V}_{\beta ,t}(x)\right] \psi _{\beta }(x-t\tau _{\beta }),
\label{pbt}
\end{equation}
where $\tau _{\beta }=\pi l(2\beta +1)$ and 
\begin{equation}
\bar{V}_{\beta ,t}(x)=\sum_{s=0}^{t-1}V\left( x-s\tau _{\beta }\right)
=\sum_{m}V_{m}\frac{\sin (m\tau _{\beta }t/2)}{\sin (m\tau _{\beta }/2)}
e^{-im[x-(t-1)\tau _{\beta }/2]}.  \label{Vbt}
\end{equation}

{\em General exact expressions.} The expectation value $\left\langle
E\right\rangle _{t}$ of the kinetic energy $\hat{p}^{2}/2$ in the wave
packet (\ref{KPR}) can be expressed in the $p$ representation with $p$
decomposed into its integer ($n$) and fractional ($\beta $) parts, $
p=n+\beta $. The resulting expression can be compactly written in terms of
the Bloch functions $\varphi _{\beta ,t}(x)=\exp (i\beta x)\psi _{\beta
,t}(x)$, using (\ref{phb}) (see note \cite{note}): 
\begin{equation}
\left\langle E\right\rangle _{t}=\frac{1}{2}\int_{0}^{1}d\beta
\sum_{n}(n+\beta )^{2}\left| \widetilde{\Psi }_{t}(n+\beta )\right| ^{2}=
\frac{1}{2}\int_{0}^{1}d\beta \int_{0}^{2\pi }dx\left| \frac{d\varphi
_{\beta ,t}(x)}{dx}\right| ^{2},  \label{ake}
\end{equation}
Under QR conditions, $\tau /(2\pi )=l$, we see that Eq. (\ref{pbt}) is
satisfied with $\varphi _{\beta }$ replacing $\psi _{\beta }$ on both sides,
up to the factor $\exp (i\beta t\tau _{\beta })$. This factor, however, will
not appear in (\ref{ake}). We then easily find that $\left\langle
E\right\rangle _{t}$ is the sum of three terms, $\left\langle E\right\rangle
_{t}=\left\langle E\right\rangle _{t}^{(1)}+\left\langle E\right\rangle
_{t}^{(2)}+\left\langle E\right\rangle _{t}^{(3)}$, where 
\begin{equation}
\left\langle E\right\rangle _{t}^{(1)}=\frac{k^{2}}{2}\int_{0}^{1}d\beta
\int_{0}^{2\pi }dx\left| \varphi _{\beta }(x)\frac{d\bar{V}_{\beta
,t}(x+t\tau _{\beta })}{dx}\right| ^{2},  \label{ake1}
\end{equation}
\begin{equation}
\left\langle E\right\rangle _{t}^{(2)}=k\int_{0}^{1}d\beta 
\ \int_{0}^{2\pi}dx {\rm Im}
\left[ \varphi _{\beta }(x)\frac{d\varphi _{\beta }^{\ast }(x)}{dx}\right] 
\frac{d\bar{V}_{\beta ,t}(x+t\tau _{\beta })}{dx},  \label{ake2}
\end{equation}
\begin{equation}
\left\langle E\right\rangle _{t}^{(3)}=\frac{1}{2}\int_{0}^{1}d\beta \
\int_{0}^{2\pi }dx\left| \frac{d\varphi _{\beta }(x)}{dx}\right| ^{2}.
\label{ake3}
\end{equation}
The constant term (\ref{ake3}) is just the initial value $\left\langle
E\right\rangle _{0}$ of $\left\langle E\right\rangle _{t}$. We now derive
more explicit expressions for (\ref{ake1}) and (\ref{ake2}). We then show
that (\ref{ake1}) is totally responsible for the asymptotic behavior of 
$\left\langle E\right\rangle _{t}$, including characteristic fluctuations.
The term (\ref{ake2}) is shown to contribute transient effects which can be
significant in some cases.\newline

To express (\ref{ake1}) more explicitly, we use an expansion following from
Eq. (\ref{phb}): 
\begin{equation}
\left| \varphi _{\beta }(x)\right| ^{2}=\left| \psi _{\beta }(x)\right| ^{2}=
\frac{1}{2\pi }\sum_{m}C_{\beta }(m)\exp (imx),  \label{pbc}
\end{equation}
where $C_{\beta }(m)$ are correlations of the KP wave packet in
angular-momentum (fixed $\beta $) space, 
\begin{equation}
C_{\beta }(m)=\sum_{n}\widetilde{\Psi }(m+n+\beta )\widetilde{\Psi }^{\ast
}(n+\beta )=\sum_{j}\bar{C}_{j}(m)\exp (2\pi ij\beta ).  \label{cbm}
\end{equation}
The last expression in (\ref{cbm}) is the Fourier expansion of $C_{\beta
}(m) $, based on the obvious periodicity $C_{\beta +1}(m)=$ $C_{\beta }(m)$.
We insert Eqs. (\ref{pbc}), (\ref{cbm}), and (\ref{Vbt}) in (\ref{ake1}) and
use the identity 
\begin{equation}
\frac{\sin (m\tau _{\beta }t/2)}{\sin (m\tau _{\beta }/2)}=\exp [im(t-1)\tau
_{\beta }/2]\sum_{s=0}^{t-1}\exp (-ims\tau _{\beta }).  \label{sint}
\end{equation}
Since $\tau _{\beta }=\pi l(2\beta +1)$, the integrations in (\ref{ake1})
can be explicitly performed and we finally obtain the general exact
expression 
\begin{equation}
\left\langle E\right\rangle _{t}^{(1)}=\frac{k^{2}}{2}\sum_{m,m^{\prime
}}mm^{\prime }V_{m}V_{m^{\prime }}^{\ast }\sum_{s,s^{\prime }=1}^{t}\left(
-1\right) ^{l(ms-m^{\prime }s^{\prime })}\bar{C}_{l(ms-m^{\prime }s^{\prime
})}(m-m^{\prime }).  \label{ake1e}
\end{equation}

Consider now the term (\ref{ake2}). Using\ $\varphi _{\beta ,t}(x)=\exp
(i\beta x)\psi _{\beta ,t}(x)$ with (\ref{phb}), we find that 
\begin{equation}
{\rm Im}
\left[ \varphi _{\beta }(x)\frac{d\varphi _{\beta }^{\ast }(x)}{dx}\right] 
=-\frac{1}{4\pi }\sum_{n,n^{\prime }}\widetilde{\Psi }(n+\beta )\widetilde{
\Psi }^{\ast }(n^{\prime }+\beta )(n+n^{\prime }+2\beta )\exp [i(n-n^{\prime
})x].  \label{Im}
\end{equation}
Let us introduce the quantity 
\begin{equation}
G_{\beta }(m)=\sum_{n}\widetilde{\Psi }(m+n+\beta )\widetilde{\Psi }^{\ast
}(n+\beta )(n+\beta )=\sum_{j}\bar{G}_{j}(m)\exp (2\pi ij\beta ),
\label{gbm}
\end{equation}
where the last expression is the Fourier expansion of $G_{\beta }(m)$, based
on the periodicity $G_{\beta +1}(m)=$ $G_{\beta }(m)$. Using Eqs. (\ref{Im}),
(\ref{Vbt}), (\ref{sint}), (\ref{cbm}), and (\ref{gbm}) in (\ref{ake2}),
we get, after a straightforward calculation, 
\begin{equation}
\left\langle E\right\rangle _{t}^{(2)}=\frac{ik}{2}\sum_{m}V_{m}
\sum_{s=1}^{t}\left( -1\right) ^{lms}\left[ m^{2}\bar{C}_{lms}(m)+2m\bar{G}
_{lms}(m)\right] .  \label{ake2e}
\end{equation}
The growth of (\ref{ake2e}) with $t$ is always slower than linear since both 
$\bar{C}_{j}(m)$ and $\bar{G}_{j}(m)$ decay with $|j|$. For sufficiently
smooth $\widetilde{\Psi }(p)$, this decay is fast enough to yield a finite
value of (\ref{ake2e}) for all $t$. Also, $\left\langle E\right\rangle
_{t}^{(2)}$ may vanish identically if some symmetry conditions are
satisfied. Let $\widetilde{\Psi }(p)$ have a definite parity, $\widetilde{
\Psi }(-p)=\pm \widetilde{\Psi }(p)$. Then, from (\ref{cbm}) and (\ref{gbm}),
$C_{\beta }(m)=C_{-\beta }(-m)$, i.e., $\bar{C}_{j}(m)=\bar{C}_{-j}(-m)$,
and $G_{\beta }(m)=-G_{-\beta }(-m)$, i.e., $\bar{G}_{j}(m)=-\bar{G}_{-j}(-m)
$. If also $V(x)$ is odd, $V(-x)=-V(x)$, i.e., $V_{-m}=-$ $V_{m}$, it
follows immediately that $\left\langle E\right\rangle _{t}^{(2)}=0$ in (\ref
{ake2e}).\newline

{\em Asymptotic time behavior.} We now show that the asymptotic behavior of 
(\ref{ake1e}) for large $t$ is typically a linear growth with bounded
fluctuations; then, since (\ref{ake2e}) grows slower than linearly (see
above), the asymptotic average growth of $\left\langle E\right\rangle _{t}$
is linear. Let us first change variables in the second sum in (\ref{ake1e})
from $(s,\ s^{\prime })$ to integer variables $(a,\ b)$ defined as follows.
We denote by $g=g(m,\ m^{\prime })$ the greatest common factor of $(|m|,\
|m^{\prime }|)$ and write $m=gm_{0}$, $m^{\prime }=gm_{0}^{\prime }$, where 
$(m_{0},\ m_{0}^{\prime })$ are coprime integers; thus, $ms-m^{\prime
}s^{\prime }=g(m_{0}s-m_{0}^{\prime }s^{\prime })$ in (\ref{ake1e}). For any
integer $a$, a solution $(s,\ s^{\prime })$ of the Diophantine equation 
$m_{0}s-m_{0}^{\prime }s^{\prime }=a$ is $s=as_{0}$ and $s^{\prime
}=as_{0}^{\prime }$, where $(s_{0},\ s_{0}^{\prime })$ are coprime integers
satisfying $m_{0}s_{0}-m_{0}^{\prime }s_{0}^{\prime }=1$; such integers
always exist for coprime $(m_{0},\ m_{0}^{\prime })$, see, e.g., Ref. 
\cite{nzm}. The general solution $(s,\ s^{\prime })$ of 
$m_{0}s-m_{0}^{\prime }s^{\prime }=a$ is then 
\begin{equation}
s=as_{0}+bm_{0}^{\prime },\ \ \ \ \ \ s^{\prime }=as_{0}^{\prime }+bm_{0}
\label{ss'}
\end{equation}
for all integers $b$. By the change of variables (\ref{ss'}), $
l(ms-m^{\prime }s^{\prime })=gla$ in (\ref{ake1e}) and, due to the decay of $
\bar{C}_{j}(m)$ with $|j|$, $\bar{C}_{gla}(m-m^{\prime })$ is negligible if $
|a|$ is too large. Now, for given $a$, the values of $b$ in (\ref{ss'}) are
restricted by the condition $1\leq s,\ s^{\prime }\leq t$ in (\ref{ake1e}).
If $(m_{0},\ m_{0}^{\prime })$ have the {\em same} sign, the number $N_{b}$
of $b$-values for $t\gg \max (|as_{0}|,|as_{0}^{\prime }|)$ is approximately 
$\left[ t/\overline{m}\right] $, where $\overline{m}=\max
(|m_{0}|,|m_{0}^{\prime }|)$ and $\left[ x\right] $ denotes the integer part
of $x$. The case of $m_{0}$ and $m_{0}^{\prime }$ having different signs can
be ignored since $N_{b}$ is much smaller than $\left[ t/\overline{m}\right] $
in this case. One can therefore approximate (\ref{ake1e}) for large enough $t
$ as follows: 
\begin{equation}
\left\langle E\right\rangle _{t}^{(1)}\approx \frac{k^{2}}{2}
\sum_{mm^{\prime }>0}mm^{\prime }V_{m}V_{m^{\prime }}^{\ast }\sum_{a}\left(
-1\right) ^{gla}\bar{C}_{gla}(m-m^{\prime })\left[ \frac{t}{\overline{m}}
\right] .  \label{ake1e2}
\end{equation}
We decompose $\left[ t/\overline{m}\right] $ as $\left[ t/\overline{m}\right]
=t/\overline{m}+F_{t}(\overline{m})$, where $F_{t}(\overline{m})=\left[ t/
\overline{m}\right] -t/\overline{m}$ are ``fluctuations'' of a simple
number-theoretical nature. Correspondingly, $\left\langle E\right\rangle
_{t}^{(1)}$ in (\ref{ake1e2}) will be decomposed into a linear-growth part $
\left\langle E\right\rangle _{t}^{(1,L)}$ and a fluctuating part $
\left\langle E\right\rangle _{t}^{(1,F)}$: $\left\langle E\right\rangle
_{t}^{(1)}\approx \left\langle E\right\rangle _{t}^{(1,L)}+\left\langle
E\right\rangle _{t}^{(1,F)}$. The expression for $\left\langle
E\right\rangle _{t}^{(1,L)}$, given by Eq. (\ref{ake1e2}) with $\left[ t/
\overline{m}\right] $ replaced by $t/\overline{m}$, can be easily shown to
be precisely the known one for the asymptotic linear growth derived in Ref. 
\cite{dd} [see Eqs. (17) and (19) there]. In fact, from Eq. (\ref{cbm}) we
find by simple algebra that 
\begin{equation}
\sum_{a}\left( -1\right) ^{gla}\bar{C}_{gla}(m-m^{\prime })=\frac{1}{gl}
\sum_{r=0}^{gl-1}C_{\beta _{r,g}}(m-m^{\prime }),  \label{cgla}
\end{equation}
where $\beta _{r,g}=r/(gl)-1/2$ ${\rm mod}(1)$, $r=0,\dots ,\ gl-1$, are
resonant values of $\beta $ \cite{dd}. Using (\ref{cgla}) and the identity $
2mm^{\prime }/(g\overline{m})=$ $|m|+|m^{\prime }|-|m-m^{\prime }|$, $
\left\langle E\right\rangle _{t}^{(1,L)}$ reduces to the expression (17)
with (19) in Ref. \cite{dd}; alternatively, $\left\langle E\right\rangle
_{t}^{(1,L)}=Dt$, where $D$ is the coefficient (20) in Ref. \cite{dd}.
\newline

The fluctuating part $\left\langle E\right\rangle _{t}^{(1,F)}$ is given by
Eq. (\ref{ake1e2}) with $\left[ t/\overline{m}\right] $ replaced by $F_{t}(
\overline{m})=\left[ t/\overline{m}\right] -t/\overline{m}$. Clearly, $
\left\langle E\right\rangle _{t}^{(1,F)}=0$ identically only if $F_{t}(
\overline{m})=0$, i.e., $\overline{m}=1$, for all relevant pairs $(m,\
m^{\prime })$. This will occur in two cases: (a) The potential is
mono-harmonic, i.e., $V_{m}=0$ for $|m|>1$. (b) The correlations $C_{\beta
}(m-m^{\prime })=0$ for $m\neq m^{\prime }$, corresponding to a uniform 
($x$-independent) distribution (\ref{pbc}). In general, using 
$|F_{t}(\overline{m})|<1$ and (\ref{cgla}), we easily see that $\left
| \left\langle E\right\rangle _{t}^{(1,F)}\right| <B=(k^{2}\tilde{C}/2)
\sum_{mm^{\prime }>0}mm^{\prime }\left| V_{m}V_{m^{\prime }}^{\ast }\right| $, 
where $\tilde{C}$ is an upper bound of $\left| C_{\beta }(m)\right| $ which 
decays with $|m|$. The quantity $B$ is always finite for a typical, 
differentiable $V(x)$ with $\left| V_{m}\right| $ decaying faster than 
$|m|^{-2}$. Thus, $\left\langle E\right\rangle _{t}^{(1,F)}$ is bounded and is 
therefore much smaller than $\left\langle E\right\rangle _{t}$ for large $t$. An 
example of significant fluctuations for small $t$ is considered below.\newline

{\em Exact closed results for a class of wave packets.} We now focus on the
class of wave packets for which $\widetilde{\Psi }(p)$ is piecewise constant
on the unit intervals $n\leq p<n+1$ for all integers $n$. This corresponds,
by Eq. (\ref{phb}), to a $\beta $-independent $\psi _{\beta }(x)$, i.e., the
same $\psi _{\beta }(x)=\psi (x)$ is associated with all the $\beta $-KRs.
For $\beta $-independent correlations in (\ref{pbc}), $C_{\beta }(m)=C(m)$,
one has $\bar{C}_{j}(m)=C(m)\delta _{j,0}$ in (\ref{cbm}). Using the last
result in the derivation above of Eq. (\ref{ake1e2}), we see that Eq. 
(\ref{ake1e2}) can be replaced by an exact formula for {\em all} times $t$: 
\begin{equation}
\left\langle E\right\rangle _{t}^{(1)}=\frac{k^{2}}{2}\sum_{mm^{\prime
}>0}mm^{\prime }V_{m}V_{m^{\prime }}^{\ast }C(m-m^{\prime })\left[ \frac{t}
{\overline{m}}\right] .  \label{ake1ee}
\end{equation}
Also, in Eq. (\ref{gbm}), $G_{\beta }(m)=G_{0}(m)+\beta C(m)$ for $0\leq
\beta <1$. Then 
\begin{equation}
\bar{G}_{j}(m)=\int_{0}^{1}d\beta \exp (-2\pi ij\beta )G_{\beta }(m)=\left[
G_{0}(m)+\frac{C(m)}{2}\right] \delta _{j,0}+i\frac{C(m)}{2\pi j}\left(
1-\delta _{j,0}\right) .  \label{gjm}
\end{equation}
Inserting $\bar{C}_{j}(m)=C(m)\delta _{j,0}$ and (\ref{gjm}) in 
(\ref{ake2e}), we obtain 
\begin{equation}
\left\langle E\right\rangle _{t}^{(2)}=-\frac{k}{2\pi l}\sum_{m\neq
0}V_{m}C(m)\sum_{s=1}^{t}\frac{(-1)^{lms}}{s}.  \label{ake2ee}
\end{equation}
Let us briefly analyze the exact results (\ref{ake1ee}) and (\ref{ake2ee}).
It is clear from the definition of $\overline{m}$ that the fluctuating part $
\left\langle E\right\rangle _{t}^{(1,F)}$ of (\ref{ake1ee}) is a periodic or
quasiperiodic function of $t$\ depending on whether the number of harmonics
of the potential is finite or infinite. Concerning (\ref{ake2ee}), we first
introduce the notations $W_{1}=-k/(2\pi l)\sum_{m}V_{2m+1}C(2m+1)$, $
W_{2}=-k/(2\pi l)\sum_{m\neq 0}V_{2m}C(2m)$, and $W=W_{1}+W_{2}$. A good
approximation of (\ref{ake2ee}) for even $l$ is then $\left\langle
E\right\rangle _{t}^{(2)}\approx W[\ln (t)+\gamma ]$, where $\gamma \approx
0.5772$ is the Euler constant. For odd $l$, $\left\langle E\right\rangle
_{t}^{(2)}\approx W_{2}[\ln (t)+\gamma ]-W_{1}\ln (2)$. Thus, (\ref{ake2ee})
provides logarithmic corrections to the dominant growth of $\left\langle
E\right\rangle _{t}$ given by (\ref{ake1ee}). These corrections vanish if,
e.g., the symmetry conditions mentioned after Eq. (\ref{ake2e}) are
satisfied. Then, $\left\langle E\right\rangle _{t}=\left\langle
E\right\rangle _{0}+$ $\left\langle E\right\rangle _{t}^{(1)}$, so that the
asymptotic behavior (\ref{ake1ee}) of $\left\langle E\right\rangle _{t}$
starts at $t=1$. If, in addition, $\left\langle E\right\rangle _{t}^{(1,F)}=0
$ (see the cases in which this occurs at the end of the previous section), $
\left\langle E\right\rangle _{t}$ is an exactly linear function of $t$. See
also note \cite{note1}.\newline

In general, $\left\langle E\right\rangle _{t}^{(2)}$ may be significant
relative to $\left\langle E\right\rangle _{t}^{(1)}$ for small $t$. As a
simple example, we consider the case of the two-harmonic potential $
V(x)=\cos (x)+\eta \cos (2x)$ with $\psi _{\beta }(x)=\psi (x)=A[1+\lambda
\cos (x)]$, where $\eta $ and $\lambda $ are constants and $A=[2\pi
(1+\lambda ^{2}/2)]^{-1/2}$ is the normalization factor, $\int_{0}^{2\pi
}dx|\psi (x)|^{2}=1$. The only nonzero values of $V_{m}$ and $C(m)$ are
those for $|m|\leq 2$ and can be easily calculated from (\ref{V}) and 
(\ref{pbc}). Inserting them in (\ref{ake1ee}) and in the results above for 
(\ref{ake2ee}), we get 
\begin{equation}
\left\langle E\right\rangle _{t}^{(1)}=k^{2}\left[ \frac{1}{4}\left( 1+\eta
^{2}\right) +\frac{\eta \lambda }{2+\lambda ^{2}}\right] t+k^{2}\frac{2\eta
\lambda }{2+\lambda ^{2}}F_{t}(2),  \label{ake1ex}
\end{equation}
\begin{equation}
\left\langle E\right\rangle _{t}^{(2)}\approx -\frac{k}{2\pi l}\frac{
2\lambda +\eta \lambda ^{2}/2}{2+\lambda ^{2}}\left[ \ln (t)+\gamma \right] ,
\label{ake2ex}
\end{equation}
where $F_{t}(2)=0$ or $-1/2$ for $t$ even or odd, respectively, and (\ref
{ake2ex}) holds for even $l$ (the case of odd $l$ can be treated similarly
and will not be discussed here). Choose $\lambda =\sqrt{2}$, so that $
|\lambda |/(2+\lambda ^{2})$ assumes its maximal value $\sqrt{2}/4$. Then,
for $\eta \approx 1$ and sufficiently small $k\ll 1$, the logarithmic
correction (\ref{ake2ex}) is significantly larger than (\ref{ake1ex}) for
all times $t$ satisfying $\ln (t)/t\gg 2\pi lk$. If $k>1$, on the other
hand, $\left\langle E\right\rangle _{t}^{(2)}$ is negligible relative to 
$\left\langle E\right\rangle _{t}^{(1)}$. For $\eta \approx 1$, the
fluctuations in (\ref{ake1ex}) have a magnitude of about $40\%$ of that of
the coefficient $D\approx \left\langle E\right\rangle _{t}^{(1)}/t$ and may
be therefore observed experimentally for not too large $t$.\newline

{\em Incoherent mixtures.} We now assume that the initial KP state is an
incoherent mixture of plane waves $\exp (ipx)$ with momentum distribution
$f(p)$ sufficiently localized in $p$. At time $t$, the expectation value of
the kinetic energy in the state evolving from a plane wave with $p=n+\beta $
is, from Ref. \cite{dd} (with a change in notation), 
\begin{equation}
E_{t}(n,\beta )=\frac{(n+\beta )^{2}}{2}+k^{2}\sum_{m>0}m^{2}|V_{m}|^{2}
\frac{\sin ^{2}(m\tau _{\beta }t/2)}{\sin ^{2}(m\tau _{\beta }/2)}.
\label{Ebt}
\end{equation}
The MKE of the mixture is then $\left\langle E_{t}\right\rangle =
\int_{0}^{1}d\beta \sum_{n}f(n+\beta )E_{t}(n,\beta )$. Defining the
function $f_{0}(\beta )=\sum_{n}f(n+\beta )$, which is periodic in $\beta $
with Fourier expansion $f_{0}(\beta )=\sum_{j}\bar{f}(j)\exp (2\pi ij\beta )$,
and using (\ref{sint}), we obtain after a straightforward calculation 
\begin{equation}
\left\langle E_{t}\right\rangle =\left\langle E_{0}\right\rangle
+k^{2}\sum_{m>0}m^{2}|V_{m}|^{2}\sum_{a=1-t}^{t-1}(-1)^{mla}\bar{f}
(mla)(t-|a|).  \label{akem}
\end{equation}
Clearly, in the asymptotic time regime $t\gg 1$, (\ref{akem}) always
exhibits a linear growth {\em without} fluctuations, unlike (\ref{ake1e2}).
For a nonuniform mixture [$f_{0}(\beta )\neq 1$, i.e., $\bar{f}(j)\neq
\delta _{j,0}$], this linear growth will be generally preceded by a
transient behavior for sufficiently small $t$.\newline

In conclusion, we have shown that for KP wave packets the asymptotic
quantum-resonant evolution of the MKE is typically given by $\left\langle
E\right\rangle _{t}\sim Dt+\left\langle E\right\rangle _{t}^{(1,F)}$, where 
$D$ is the coefficient of linear growth \cite{dd} and $\left\langle
E\right\rangle _{t}^{(1,F)}$ are bounded fluctuations of a simple
number-theoretical nature. For a large class of initial wave packets,
$\left\langle E\right\rangle _{t}$ is exactly the superposition of its
asymptotic behavior and transient logarithmic corrections which may be
negligible for large $k$ or can vanish under some conditions. The asymptotic
behavior then starts already at small $t$, where the fluctuations are most
prominent, especially when the magnitude of $\left\langle E\right\rangle
_{t}^{(1,F)}$ is comparable to that of $D$ as in the example above. We
remark here that the sensitive dependence of $D$ on the harmonics of $V(x)$
and on the initial wave packet, pointed out in Ref. \cite{dd}, is featured
also by $\left\langle E\right\rangle _{t}^{(1,F)}$, as one can clearly see
from (\ref{ake1e2}) or (\ref{ake1ex}). The fluctuations in $\left\langle
E\right\rangle _{t}$ for small $t$ may be noticeable even when the transient
corrections are significant. All the quantum-resonance behaviors for $t$ not
too large are expected to be robust under sufficiently small deviations of
$\tau /(2\pi )$ from integers and may be therefore observed experimentally.
\newline

Since integer values of $\tau /(2\pi )$ correspond to strong quantum
regimes, the fluctuations and transient phenomena studied here are basically
different in nature from known ones occurring in semiclassical regimes 
\cite{qc,gwb,dd1}, e.g., the asymptotic quasiperiodic fluctuations associated
with dynamical localization and transient behaviors featuring an approximate
coincidence of the classical and quantum evolutions on some initial time
interval. It would be interesting, however, if one could establish some
correspondence between the general quantum-resonance phenomena in this paper
and classical behaviors of related systems, as it was done in a particular
case in Ref. \cite{kp2}.\newline

This work was partially supported by the Israel Science Foundation (Grant
No. 118/05). DLD acknowledges partial support from the Russian Ministry of
Education and Science and the US Civilian Research and Development
Foundation (CRDF BRHE Program, Grants Nos. VZ-0-010 and Y2-P-10-01).\newline


\begin{references}
\bibitem{note1}  For odd $l$, the logarithmic corrections will vanish also
if $W_{2}=0$, e.g., when $V(x)$ contains only odd harmonics. Then,
$\left\langle E\right\rangle _{t}^{(2)}\approx -W_{1}\ln (2)$ and, if
$\left\langle E\right\rangle _{t}^{(1,F)}=0$, $\left\langle E\right\rangle
_{t}$ will be very close to a linear function of $t$.
\end{references}
\end{document}